\documentclass[aps, pra, twocolumn, superscriptaddress]{revtex4-2}

\usepackage[pdftex]{graphicx}
\usepackage{amsmath}
\usepackage{amssymb}
\usepackage{physics}
\usepackage{nicefrac}
\usepackage{siunitx}
\usepackage{bbold}
\usepackage{braket}
\usepackage{ragged2e}
\usepackage{lipsum}
\usepackage{multirow}
\setcitestyle{numbers}
\usepackage{subcaption}
\usepackage{layouts}
\usepackage{framed}
\usepackage{float}
\usepackage{bm}
\usepackage{mathtools}

\usepackage[dvipsnames]{xcolor}
\usepackage{hyperref}
\hypersetup{
  colorlinks   = true, 
  urlcolor     = MidnightBlue, 
  linkcolor    = MidnightBlue, 
  citecolor   = Maroon 
}

\renewcommand{\arraystretch}{2.0}

\begin{document}

\title{Measurement-free, scalable and fault-tolerant universal quantum computing}

\author{Friederike Butt}
\email[Email to ]{friederike.butt@rwth-aachen.de}
\affiliation{Institute for Quantum Information, RWTH Aachen University, Aachen, Germany}
\affiliation{Institute for Theoretical Nanoelectronics (PGI-2), Forschungszentrum J\"{u}lich, J\"{u}lich, Germany}

\author{David F. Locher}
\affiliation{Institute for Quantum Information, RWTH Aachen University, Aachen, Germany}
\affiliation{Institute for Theoretical Nanoelectronics (PGI-2), Forschungszentrum J\"{u}lich, J\"{u}lich, Germany}

\author{Katharina Brechtelsbauer}
\affiliation{Institute for Theoretical Physics III and Center for Integrated Quantum Science and Technology, University of Stuttgart, Stuttgart, Germany}

\author{Hans Peter B\"uchler}
\affiliation{Institute for Theoretical Physics III and Center for Integrated Quantum Science and Technology, University of  Stuttgart, Stuttgart, Germany}

\author{Markus M\"{u}ller}
\affiliation{Institute for Quantum Information, RWTH Aachen University, Aachen, Germany}
\affiliation{Institute for Theoretical Nanoelectronics (PGI-2), Forschungszentrum J\"{u}lich, J\"{u}lich, Germany}

\date{\today}

\begin{abstract}

Reliable execution of large-scale quantum algorithms requires robust underlying operations and this challenge is addressed by quantum error correction (QEC). 
Most modern QEC protocols rely on measurements and feed-forward operations, which are experimentally demanding, and often slow and prone to high error rates. 
Additionally, no single error-correcting code intrinsically supports the full set of logical operations required for universal quantum computing, resulting in an increased operational overhead. In this work, we present a complete toolbox for fault-tolerant universal quantum computing without the need for measurements during algorithm execution by combining the strategies of code switching and concatenation. To this end, we develop new fault-tolerant, measurement-free protocols to transfer encoded information between 2D and 3D color codes, which offer complementary and in combination universal sets of robust logical gates. We identify experimentally realistic regimes where these protocols surpass state-of-the-art measurement-based approaches. 
Moreover, we extend the scheme to higher-distance codes by concatenating the 2D color code with itself and by integrating code switching for operations that lack a natively fault-tolerant implementation. 
Our measurement-free approach thereby provides a practical and scalable pathway for universal quantum computing on state-of-the-art quantum processors. 

\end{abstract}

\maketitle

\section{Introduction}

\begin{figure*}[!tb]
	\centering
	\includegraphics[width=\linewidth]{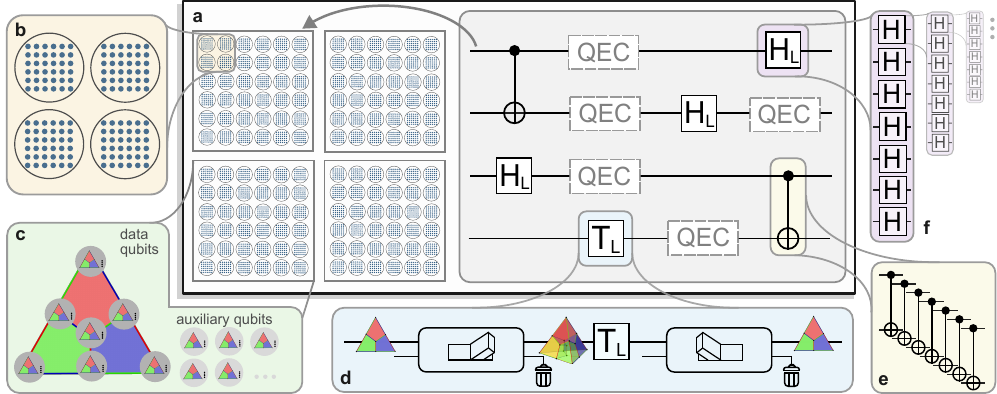}
 \caption{\justifying \textbf{Measurement-free universal quantum computing by means of concatenation and code switching.} \textbf{a}~A quantum algorithm can be constructed from a universal logical gate set such as \{H$_L$, CNOT$_L$, T$_L$\} and QEC to maintain fault tolerance in an algorithm. We provide a complete toolbox to run these circuits fault-tolerantly on logical qubits, which are encoded in blocks of physical qubits. \textbf{b}~Physical qubits form logical qubits, which in turn again encode logical qubits, a scheme known as concatenation. \textbf{c}~We choose the seven-qubit color code as the base-code of our protocols, which is concatenated with itself and requires a set of auxiliary qubits. \textbf{d}~To apply the logical gate T$_L$, we switch to a 3D color code that has a transversal implementation. Afterwards, the encoded quantum information is transferred back to the initial code. \textbf{e, f}~The Clifford operations H$_L$ and CNOT$_L$ can be performed transversally on the $[[7, 1, 3]]$ code and thus in a natively FT way by bitwise application of the respective physical operations.}
	\label{fig:Fig_1_motivation}
\end{figure*}

A key requirement for the practical deployment of quantum algorithms is their robustness against noise, alongside the capacity to implement arbitrary operations on qubits. 
Quantum error correction (QEC) provides protection against noise by enabling the detection and correction of errors that arise during computation~\cite{Nielsen_and_Chuang}, and recent experiments have demonstrated significant breakthroughs in the field~\cite{putterman2024hardware, ryan2024high, acharya2024quantum, bluvstein2024logical}.
The latter is realized through a discrete, universal set of gates capable of approximating any quantum operation to in principle arbitrary precision~\cite{Nielsen_and_Chuang}. 
Fault-tolerant (FT) implementations of these gates prevent the uncontrolled propagation of errors through suitable quantum circuit design principles~\cite{knill1998resilient}. However, achieving such FT implementations of a full universal gate set poses a significant challenge, as no known QEC code intrinsically supports a fully FT universal gate set~\cite{eastin2009restrictions}. 
Two well-established methods to complete a FT universal gate set are magic state injection and code switching. Magic state injection makes use of a fault-tolerantly prepared logical magic resource state~\cite{goto2016minimizing} which is injected onto the encoded data qubit~\cite{bravyi2005universal}. 
Code switching enables the combination of two codes with complementary sets of transversal gates by transferring encoded information between them~\cite{bombin2016dimensional, anderson2014fault}. 
Recent experiments have demonstrated a FT universal gate set by means of code switching for the first time~\cite{pogorelov2024experimental} and first FT computations in combination with error correction~\cite{reichardt2024demonstration, ryan2024high, bluvstein2024logical, nguyen2021demonstration}.
However, the success probability of many practical protocols is fundamentally limited by mid-circuit measurements, which is challenging on many hardware platforms. For instance, in atomic setups, such as trapped ions and neutral atoms, fluorescence measurements heat up the atoms which requires additional laser cooling during or after the read-out. 
Moreover, in atomic as well as superconducting quantum processors, measurements are still orders of magnitude slower than typical gate times which leads to decoherence of idling qubits in the meanwhile and imposes severe speed limitations~\cite{pogorelov2024experimental, postler2023demonstration, moses2023race, graham2023midcircuit, acharya2024quantum}.
Real-time decoding~\cite{ryan2024high, acharya2024quantum} and feedback based on measurement outcomes has been realized but is still experimentally demanding~\cite{singh2023mid, ryan2024high}. In contrast to this, resetting qubits can typically be done fast non-destructively. 
These limitations and experimental capabilities motivate the search for FT protocols which do not rely on mid-circuit measurements or feed-forward operations. Recently, \textit{measurement-free} (MF) schemes for state preparation~\cite{goto2023measurement} and QEC on different codes have been constructed~\cite{perlin2023fault, heussen2024measurement, veroni2024optimized}. 
The idea behind these MF QEC schemes is to transfer the stabilizer information onto additional auxiliary qubits, and perform decoding as well as coherent feedback within the quantum algorithm itself. 
Finally, auxiliary qubits can be reset to be reused or substituted with fresh qubits, which effectively removes the entropy introduced by the noise. However, so far an approach to implement a FT universal gate set without relying on measurements is missing, and furthermore, the existing MF schemes only consider small, low-distance code instances, and do not provide a method to scale this approach to larger distance
codes with increased protection. 

In this work, we show how quantum computers can be run autonomously, without measurement interventions, freely programmable and yet in a fault-tolerant manner. We achieve this by developing a scheme for freely scalable fault-tolerant (FT) and measurement-free (MF) quantum computing that combines code switching and code concatenation. First, we construct new MF FT code switching schemes to transfer encoded information between the smallest instances of a 2- and a 3-dimensional color code. This enables the implementation of a deterministic FT universal gate set which does not require measurements or feed-forward operations during the execution of a logical quantum algorithm. Then, we scale our schemes to high distances by concatenating a code block with itself and including switches, as illustrated in Fig.~\ref{fig:Fig_1_motivation}.

\section{Measurement-free fault-tolerant code switching}

\begin{figure*}[t]
	\centering
	\includegraphics[width=\linewidth]{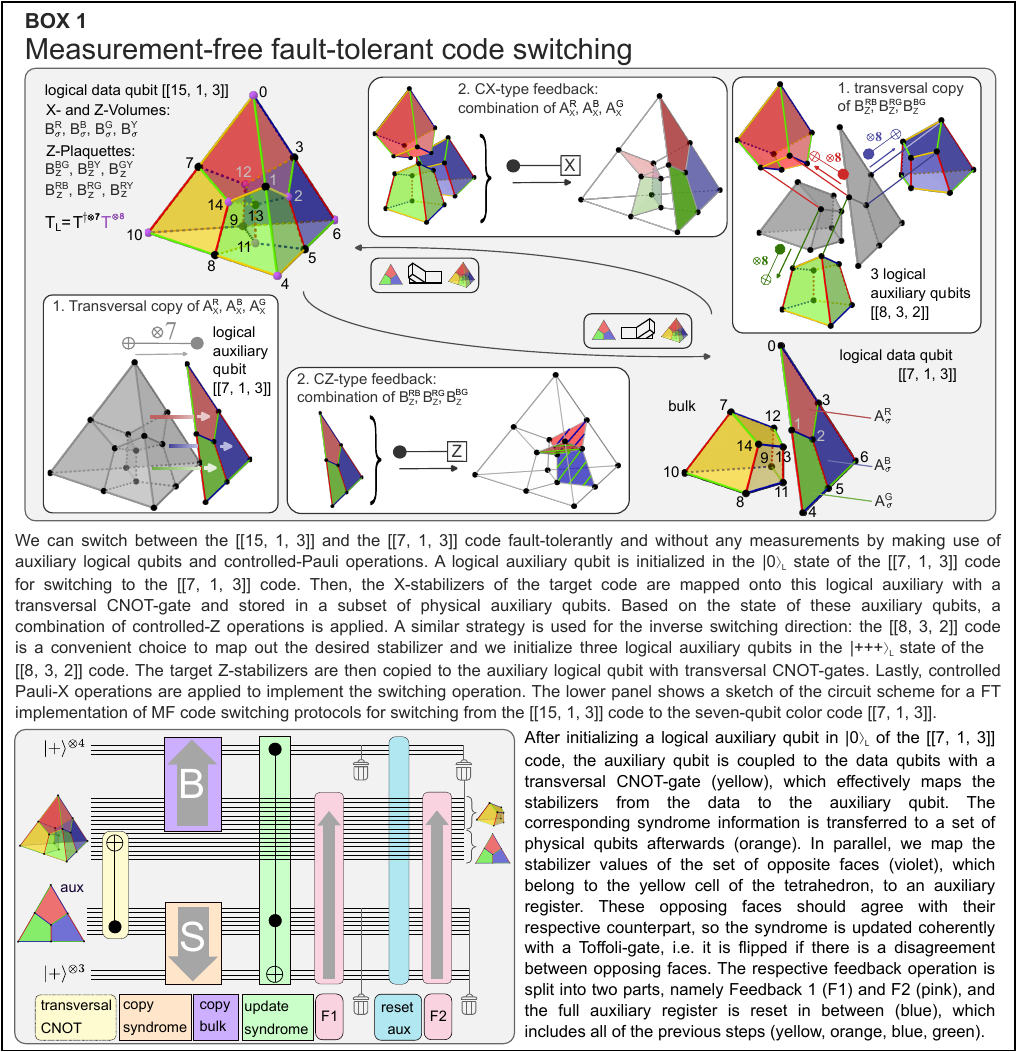}
	\label{fig:box_1}
\end{figure*}

The smallest instance of the 2D color code \mbox{$[[n = 7, k = 1, d = 3]]$} encodes a single logical qubit $k = 1$ in seven physical qubits $n = 7$ and has distance $d = 3$, meaning that any single error can be corrected~\cite{steane1996multiple}. 
Three X- and Z-stabilizers are defined symmetrically on the plaquettes formed by four physical qubits, as illustrated in Box~\hyperref[fig:box_1]{1} and Methods Fig.~\ref{fig:codes_definition}a. The logical Pauli-operators correspond to applying X- and Z-operations to all seven qubits and a logical Hadamard operation can be implemented transversally by applying seven single-qubit Hadamard gates. In 3D, the smallest error-correcting instance is the tetrahedral \mbox{$[[n = 15, k = 1, d = 3]]$} color code, which encodes $k = 1$ logical in $n = 15$ physical qubits and has distance $d = 3$~\cite{bombin2007topological}. The 
stabilizer generators of this code are given in Box~\hyperref[fig:box_1]{1} and Methods Fig.~\ref{fig:codes_definition}b, and the logical X- and Z-operators of this code coincide with those of the seven-qubit color code. The X- and Z-stabilizers are defined on different support, thus not allowing the transversal Hadamard gate. However, a transversal FT non-Clifford T-gate can be implemented by applying physical T- and T$^{\dag}$-operations in a pre-defined pattern to all fifteen qubits. The combination of these fully transversal gates, together with FT code switching gives rise to a fully transversal universal gate set. 
We first review the existing code switching procedure, before discussing the extension to a \textit{measurement-free} setting. \\

\noindent \textbf{Measurement-based code switching}\\
With measurement-based code switching~\cite{pogorelov2024experimental, butt2023fault, anderson2014fault, heussen2024efficient}, one can transfer encoded information between the two codes introduced above by, first, measuring the subset of stabilizers of the target code which are not shared with the initial code. This randomly initializes the state in a $+1$- or $-1$-eigenstate of the measured operators. Second, local Pauli-generators are applied to bring the state into a $+1$-eigenstate of all target stabilizers without changing the logically encoded information~\cite{anderson2014fault, bombin2016dimensional}. 
Specifically for switching from $[[15, 1, 3]]$ to $[[7, 1, 3]]$, we measure the three X-stabilizers $(A_X^R, A_X^B, A_X^G)$ of $[[7, 1, 3]]$ as shown in Box~\hyperref[fig:box_1]{1}. The initial tetrahedral code is a $+1$-eigenstate of the weight-8 cells, so this measurement yields random outcomes $\pm1$ for each target stabilizer. Next, a combination of the Z-stabilizers connecting the $[[7, 1, 3]]$-instance with the yellow cell is applied, i.e.~a combination of $(B_Z^{RB}, B_Z^{RG}, B_Z^{BG})$. For example, in the first step we could find the outcome $(A_X^R, A_X^B, A_X^G) = (0, 1, 0)$ and, based on this measurement, then apply $B_Z^{RG}$%$ = Z_1 Z_2 Z_{13} Z_{14}$. 
, which shares an even number of sites with the red and green plaquette and a single site with the blue one, thus fixing the state into the desired target codespace. 
This procedure is inverted for switching from $[[7, 1, 3]]$ to $[[15, 1, 3]]$: First the Z-stabilizers connecting the 2D color code instance with the yellow cell are measured, i.e.~$(B_Z^{RB}, B_Z^{RG}, B_Z^{BG})$, and a combination of $(A_X^R, A_X^B, A_X^G)$ is applied. The respective quantum feedback obeys a lookup-table-like logic as summarized in Methods Sec.~\ref{app:LUT_switching}. \\

\begin{figure*}[!tb]
	\centering
	\includegraphics[width=\linewidth]{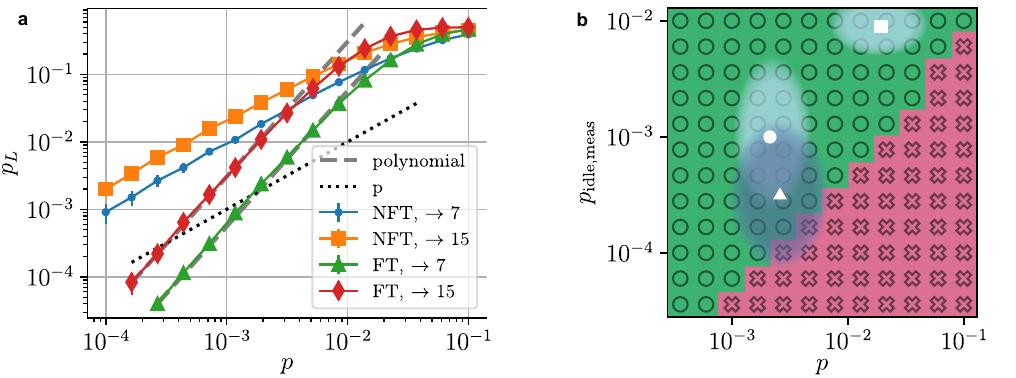}
	\caption{\justifying \textbf{MF FT code switching and logical operations. }\textbf{a} Logical failure rates for switching from the $[[15, 1, 3]]$ to the $[[7, 1, 3]]$ code and the inverse direction for the FT and non-FT MF protocol, averaged over different logical input states. All two-qubit gate components infer an error with the error rate $p$ and any single-qubit gate operation introduces an error with a probability $\frac{p}{10}$. The grey dashed lines correspond to the approximated polynomial (Eq.~\eqref{eq:error_polynomial}) with the coefficients given in Methods Tab.~\ref{tab:coefficients_FP_counting}. The black dotted line is the physical error rate, indicating a break-even point for the FT schemes at $p \approx 3\cdot 10^{-4}$. \textbf{b} Difference in the logical error rate between the measurement-based and the MF protocol for the FT H$_L$-gate on the tetrahedral $[[15, 1, 3]]$ code. The MF protocol achieves lower logical failure rates in the area depicted in green with circles, while the measurement-based version yields lower logical failure rates in the the area shown in red with crosses. The symbols indicate parameter regimes demonstrated in experiments with trapped ions in static traps $(\square)$~\cite{pogorelov2024experimental, postler2023demonstration}, shuttling-based traps $(\bigcirc)$~\cite{moses2023race}, and with neutral atoms in tweezer arrays $(\triangle)$~\cite{evered2023high, bluvstein2024logical, tsai2024benchmarking, radnaev2024universal}
	\label{fig:results_main}. }
\end{figure*}

\noindent \textbf{Measurement-free fault-tolerant code switching}\\
The main idea behind MF code switching is to map the desired stabilizer information to an auxiliary register, but instead of measuring and providing classical feedback, a quantum feedback operation is directly applied with controlled-Pauli-gates as part of the quantum algorithm itself. The entropy is then removed by resetting the auxiliary qubits or replacing them with fresh auxiliary qubits. In the following, we translate the above scheme for FT code switching to a \textit{measurement-free} setting, which poses several challenges: (i) the stabilizer information has to be coherently transferred to the auxiliary register in a reliable way, (ii) the randomly initialized stabilizer value has to be distinguished from a single error that flipped a given syndrome bit, and (iii) the coherent feedback operation has to be FT.

The first challenge can be resolved by using suitable logical auxiliary qubits. The set of target stabilizers can then be mapped to these logical auxiliary qubits with purely transversal operations, and subsequently stored on a subset of physical qubits, as indicated in Box~\hyperref[fig:box_1]{1}. 
The logical auxiliary qubit has to be chosen such that it shares specific stabilizers with the initial and the target code. It has to be a $+1$-eigenstate of the target stabilizers to ascertain the desired stabilizer values. Additionally, the logical auxiliary qubit has to share the respective complementary Pauli-type stabilizers of the data qubit to avoid unwanted back-propagation of Pauli-operators onto the data qubits. 
Here, we use the logical $|0\rangle_L$ of the seven-qubit color code for switching from $[[15, 1, 3]]$ to $[[7, 1, 3]]$. This code shares the three X-stabilizers with the target code as well as the Z-plaquettes with the initial tetrahedral code and is therefore a suitable candidate for MF code switching. 
Analogously, we use three $[[8, 3, 2]]$ code instances in the $|+$$++\rangle_L$ state, as defined in Methods Fig.~\ref{fig:codes_definition}c, for the inverse switching direction enabling the reliable copying of all desired stabilizer operators, which is discussed further in Methods Sec.~\ref{app:FT_MF_CS}. Both of these auxiliary codes have to be initialized in a MF manner themselves. We build on circuits for the logical $[[7, 1, 3]]$ code, developed in~\cite{heussen2024measurement, veroni2024optimized}, and construct new circuits for the MF initialization of the $[[8, 3, 2]]$ auxiliary qubits, given in Methods Sec.~\ref{app:MF_logical_init}. 

The second challenge is to identify if an error on a data qubit has propagated onto the auxiliary register and inverted the extracted stabilizer value. 
Without any additional information, it is not possible to identify these errors, since the state is initialized randomly in a $+1$ or $-1$-eigenstate of the stabilizer. 
However, we can distinguish the randomly initialized stabilizer value from these potential errors by comparing pairs of opposing faces belonging to the same cell, which should agree in the fault-free case~\cite{poulin2005stabilizer}. 
The syndrome is coherently updated with Toffoli-gates which flips the respective syndrome bit if two opposing faces disagree. 

Finally, the coherent quantum feedback operation has to be applied to a set of data qubits. 
State-of-the-art MF protocols for QEC~\cite{veroni2024optimized, heussen2024measurement, perlin2023fault, crow2016improved, paz2010fault} rely on multi-qubit Toffoli-gates to implement a lookup table feedback operation for small codes. But in contrast to QEC, we can implement this feedback operation in an iterative manner, only relying on two-qubit controlled-Pauli-operations, which is discussed further in Methods Sec.~\ref{app:LUT_switching}.
These switching operations correspond to multiple successive two-qubit gates with the same auxiliary control-qubit but different data target-qubits. 
The overall feedback operation is split into several parts in order to achieve fault tolerance. In between, we reset the syndrome and repeat the previous steps for coherent syndrome extraction. 
Otherwise, a single fault on a qubit storing one of the syndrome bits would propagate onto all four participating qubits and result in a logical failure. Box~\hyperref[fig:box_1]{1} illustrates the MF FT switching procedure and summarizes the protocol for switching from $[[15, 1, 3]]$ to $[[7, 1, 3]]$. The scheme for MF FT switching in the inverse direction is constructed conceptually analogously and is discussed further in Methods Sec.~\ref{app:FT_MF_CS}, and all explicit circuits are shown in Methods Sec.~\ref{app:resources_and_circuits}. \\

\noindent \textbf{Results}\\
We perform Monte Carlo simulations~\cite{pecos_git} and implement circuit level noise, as specified in Methods Sec.~\ref{app:numerical_methods}. Here, we focus first on a single-parameter noise model, where every two-qubit gate in the circuit introduces an error with probability $p$ and each single-qubit operation is faulty with probability $\frac{p}{10}$. 
In this setting, we find that the FT schemes outperform their non-FT counterpart below physical error rates of $p \approx 2\cdot 10^{-2}$ and $p \approx 10^{-2}$, in the two switching directions respectively, as shown in Fig.~\ref{fig:results_main}a. 
We estimate the performance of a logical gate, which does not have a transversal implementation on the given code, by simulating a full cycle of switching back and forth and extract a break-even point as shown in Methods Fig.~\ref{fig:methods_full_switching_cycle} at $p_{\mathrm{th}} \approx 2.6\cdot 10^{-4}$. 
Furthermore, we compare the performance of the FT MF logical gate to a FT measurement-based version of this protocol~\cite{butt2023fault} and determine which scheme achieves lower logical failure rates. To this end, we introduce an additional parameter $p_{\mathrm{idle, meas}}$, which indicates the error rate on idling data qubits during measurements. We identify a large parameter regime where the MF logical operation outperforms the measurement-based version, shown in green in Fig.~\ref{fig:results_main}b. 

\section{Scalability of MF FT universal gates}\label{sec:scalability}

\begin{figure*}[!tb]
	\centering
	\includegraphics[width=\linewidth]{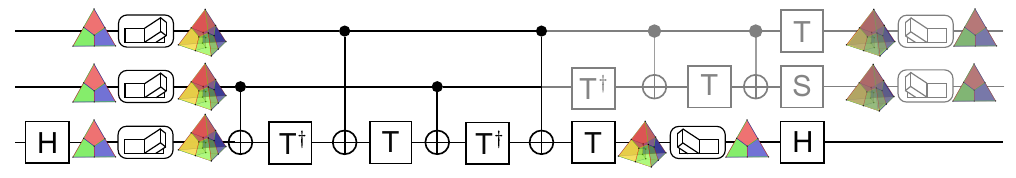}
	\caption{\justifying \textbf{ Logical operations on the concatenated $[[7, 1, 3]]$ code. } Decomposition of a Toffoli-gate into single- and two-qubit gates. If the control-qubits are reset after this gate, we do not need to execute the operations shown in grey. We additionally need to switch from the $[[7, 1, 3]]$ to the tetrahedral $[[15, 1, 3]]$ code before applying the non-Clifford T-gate in the concatenated regime where each line corresponds to a logical $[[7, 1, 3]]$ color code qubit. }
	\label{fig:results_concatenated_code}
\end{figure*}

Scaling up a FT quantum computing architecture to high-distance codes is a crucial step towards building practical, large-scale quantum computers which require low error rates~\cite{oskin2002practical}. However, this task presents major challenges in the measurement-based setting, as qubit overhead, computational complexity and hardware requirements increase significantly. Code concatenation offers a powerful method for constructing high-distance codes from smaller ones, since the failure probability is suppressed doubly-exponentially below the threshold with each layer of concatenation while maintaining polynomial-time decoding complexity~\cite{knill1998resilient, aliferis2005quantum}. 
In the following, we scale the presented MF FT protocols to high-distance codes by concatenating the $[[7, 1, 3]]$ color code with itself and combining this with code switching, thus effectively giving access to the logical T-gate for the concatenated $[[7, 1, 3]]$ code.

As a first step, we estimate the leading order contributions to the logical failure rate for the initial, non-concatenated code. The noisy two-qubit gates and the three-qubit Toffoli-gates dominate the total logical failure rate, since the number of two-qubit gates is orders of magnitude larger than that of single-qubit operations. This allows us to approximate the effective error polynomial, as discussed further in Methods Sec.~\ref{app:effective_polynomial}, as
\begin{align}
    p_L \approx c_2 p_2^2 +  c_{2, \mathrm{toff}} p_2 p_{\mathrm{toff}} + c_{\mathrm{toff}} p_{\mathrm{toff}}^2 + \mathcal{O}(p^3). 
    \label{eq:error_polynomial}
\end{align}
The coefficients $c_2, c_{2, \mathrm{toff}}$ and $c_{\mathrm{toff}}$ correspond to the number of weight-2 faults on the specified components (two-qubit gates, Toffoli-gates) that lead to a logical failure and $p_2, p_{\mathrm{toff}}$ are the error rates on the respective circuit component. 
We determine the coefficients $c_2$ and $c_{2, \mathrm{toff}}$ for the different protocols by deterministically placing all weight-2 fault configurations and counting the number of faults leading to a failure, as summarized in Methods Tab.~\ref{tab:coefficients_FP_counting}. 
We find that the approximated polynomial Eq.~\eqref{eq:error_polynomial} indeed fits the logical failure rates, as shown in Fig.~\ref{fig:results_main}a.

Now, we concatenate our scheme with the seven-qubit color code by replacing each physical qubit with another encoded $[[7, 1, 3]]$ code. Each operation in our previous circuit then corresponds to a logical operation, as for example a physical CNOT-gate is translated into a transversal two-qubit gate CNOT$^{\otimes 7}$. Analogously, each physical error rate in Eq.~\eqref{eq:error_polynomial} is replaced by the respective logical gate error rate,~e.g. the physical two-qubit gate error rate $p_2$ is replaced by the transversal CNOT gate error rate $p_2^{(1)}$, where the superscript $(1)$ indicates the level of concatenation. However, the Toffoli-gate cannot be simply translated in the same way to the next concatenation level since it involves non-Clifford operations, which are not natively FT on the seven-qubit color code. The concatenated Toffoli-gate can be realized by including code switching steps, as illustrated in Fig.~\ref{fig:results_concatenated_code}. 

Continuing to concatenate with the $[[7, 1, 3]]$ code up to concatenation level $l+1$, we find that the noisiest components of the level~$l+1$ logical T-gate are the Toffoli-gates. These Toffoli-gates are themselves dominated by code switching steps on concatenation level~$l$ and, thus, the Toffoli-gates of level~$l$ and so on. 
This in turn means, that the pseudothreshold of the logical operations of level~$l>1$ is approximately given by the pseudothreshold of the concatenated Toffoli-gate of level~$l \geq 1$, as shown in Methods Fig.~\ref{fig:methods_full_switching_cycle}. 
It is shifted to a slightly lower value as compared to that of the logical T$_L$ of level~1, since the logical Toffoli-gate now contains multiple code switching steps. 
However, we can estimate a lower bound of the pseudothreshold of the concatenated logical gate of level~$l > 1$ of $p_{\mathrm{th}}^{l > 1} \approx 1\cdot 10^{-4}$, based on the approximated effective error polynomials summarized in Methods Tab.~\ref{tab:summary_concatenated_rates}. \\

In summary, we have constructed a toolbox for implementing any single logical operation fault-tolerantly and measurement-free, which is scalable to larger code distances by concatenating the seven-qubit color code with itself and introducing the MF FT switch to realize logical Toffoli-gates. However, running algorithms fault-tolerantly requires an additional building block, namely QEC~\cite{gottesman2013fault}. 
Previous analyses on how to integrate QEC into a quantum algorithm~\cite{aliferis2005quantum, cross2007comparative, gutierrez2015comparison} have been extended to concatenated codes~\cite{chamberland2016thresholds, pato2024concatenated} and recent works have shown that $\mathcal{O}(1)$ rounds of stabilizer extraction for each logical operations can be sufficient for specific FT quantum algorithms~\cite{zhou2024algorithmic}. 

Recent schemes for FT \textit{measurement-free} QEC~\cite{veroni2024optimized, heussen2024measurement, perlin2023fault} rely on Toffoli-gates to implement corrections according to a lookup table based quantum feedback for the $[[7, 1, 3]]$ code. We can integrate this scheme into our framework by concatenating the code with itself and including the MF FT switch for each logical Toffoli-gate. Note that the \textit{measurement-free} QEC cycles coherently implement quantum feedback without the need of classical information processing. 
The cost of one QEC round below the pseudothreshold for concatenation level~$l>1$ is much smaller than the cost of a logical T-gate on the concatenated $[[7, 1, 3]]$ code, as discussed in Methods Sec.~\ref{app:cost_of_QEC}. Therefore, the pseudothreshold of the combined block of this logical operation followed by QEC is still approximately given by the pseudothreshold of the bare logical operation. 

\section{Outlook}

The presented \textit{measurement-free} and fault-tolerant implementation of a universal gate set provides a route towards scalable fault-tolerant quantum computing. Previous works on universal gate sets by means of concatenated codes~\cite{chamberland2016thresholds, jochym2014using, yoder2016universal} solely rely on the concatenation of different code types with complementary sets of gates. These require at least 49 and up to 105 physical qubits to realize a universal gate set for a distance-3 code, while in our approach, 35 qubits are sufficient, provided qubit reset is available, which reduces the experimental requirements significantly. Remarkably, the pseudothresholds of our protocols are competitive and lie in between the lower and upper bound indicated in these works~\cite{chamberland2016thresholds}. 

Our schemes provide a feasible and scalable approach for MF FT universal quantum computing. They are built on heavily parallelizable physical operations which can be implemented efficiently in experimental platforms that offer long-range connectivity between qubits.
Neutral atom platforms, for example, have demonstrated massively parallelized Clifford operations~\cite{evered2023high} as well as shuttling of entire logical qubits~\cite{bluvstein2024logical}, which are key building blocks of our protocols. Furthermore, mid-circuit measurements and real-time feedback are still experimentally demanding due to relatively slow measurements, while single- and two-qubit gate fidelities are high~\cite{evered2023high, bluvstein2024logical, radnaev2024universal, tsai2024benchmarking}. These features make neutral atom platforms an ideal candidate for MF protocols and concatenated code constructions~\cite{xu2024constant}. 
Complementary to neutral atom platforms, also trapped-ion quantum processors have demonstrated the capabilities required for handling concatenated codes in 2D architectures~\cite{hakelberg2019interference, valentini2024demonstration} and shuttling based approaches are in principle able to host the presented code constructions~\cite{reichardt2024demonstration, pino2021demonstration, moses2023race, jain2024penning}.
Also superconducting platforms with long-range couplers are advancing towards the realization of non-local connectivity~\cite{marxer2023long}, while spin-qubit quantum computing architectures have shown progress along these lines, leveraging shuttling-based techniques~\cite{kunne2024spinbus, de2024high}. 

Tailoring of the theoretical proposal to a given experimental platform is expected to further increase the logical success rates. This includes the adaptation to a biased noise setting which is present in various experimental architectures~\cite{pogorelov2024experimental, bluvstein2024logical, radnaev2024universal} and might simplify the presented protocols, significantly reducing the hardware requirements. The implementation of natively supported multi-qubit gates~\cite{evered2023high} 
could further mitigate hardware limitations. 
Additionally optimizing the integration of QEC into a logical algorithm~\cite{chamberland2016thresholds, zhou2024algorithmic} by investigating how often and on which concatenation levels QEC should be carried out offers the potential for improved logical error rates. 
Overall, our findings outline a practical pathway toward fully scalable fault-tolerant quantum computing, leveraging a completely measurement-free approach that makes our method feasible for various experimental state-of-the-art quantum hardware platforms.

\bibliography{references}

\clearpage
\section*{Methods}

\setcounter{section}{0}

\section{Fault-tolerant code switching}\label{app:FT_MF_CS}
\begin{figure*}[!tb]
	\centering
	\includegraphics[width=0.66\linewidth]{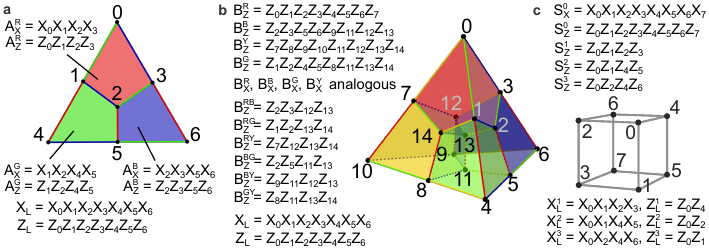}
	\caption{\justifying \textbf{Stabilizer operators of codes used for MF FT code switching} \textbf{a} The X- and Z-stabilizers of the $[[7, 1, 3]]$ color code are defined symmetrically on the red, blue and green plaquette~\cite{ steane1996multiple}. \textbf{b} Four X-stabilizers of the $[[15, 1, 3]]$ code have support on the eight qubits which form one cell. Four Z-stabilizers are defined analogously on these four cells as well as six additional independent Z-stabilizers on the weight-4 interfaces between cells~\cite{bombin2007topological, kubica2015universal}. \textbf{c} One X- and Z-stabilizer of the $[[8, 3, 2]]$ code each has support on all eight qubits. Three additional Z-stabilizers are defined on the faces of the cube~\cite{wang2024fault}. }
	\label{fig:codes_definition}
\end{figure*}

It is possible to transfer encoded information between the $[[7, 1, 3]]$ and the $[[15, 1, 3]]$ code because they correspond to two gauges of the same subsystem code~\cite{poulin2005stabilizer}. 
A subsystem code is defined by its gauge group $\mathcal{G}$, which describes a general subgroup of the $n$-qubit Pauli group~\cite{poulin2005stabilizer, kribs2005unified, butt2023fault}. The gauge group $\mathcal{G}$ of the tetrahedral subsystem code is generated by all independent X- and Z-type faces of the tetrahedral structure shown in Fig.~\ref{fig:codes_definition}b. 
The subsystem's stabilizer group $\mathcal{S} \subseteq \mathcal{G}$ is the center of $\mathcal{G}$ and it is generated by those elements commuting with all other elements in $\mathcal{G}$, which are the weight-8 cells $B_{\sigma}^R, B_{\sigma}^B, B_{\sigma}^G, B_{\sigma}^Y$ with $\sigma = $X, Z. Compared to the stabilizers of the tetrahedral $[[15, 1, 3]]$ stabilizer code as defined in Fig.~\ref{fig:codes_definition}b, the Z-stabilizers of the subsystem code are not defined on the ten independent faces of the code, but only on the four weight-$8$ cells. On the tetrahedral $[[15, 1, 3]]$ stabilizer code, the gauge of the subsystem is therefore fixed such that the codestate is not only a $+1$-eigenstate of the weight-8 cells but also of the Z-faces within the tetrahedron. In the regime of the seven-qubit color code $[[7, 1, 3]]$, in addition to the cells also the three weight-4 X- and Z-faces, as shown in Fig.~\ref{fig:codes_definition}a, are fulfilled. 

Fig.~\ref{fig:high_level_circuit_scheme_7_to_15} shows the high-level circuit scheme for MF FT switching from the 2D $[[7, 1, 3]]$ code to the 3D $[[15, 1, 3]]$ code. Analogously to the inverse direction discussed in Box~\hyperref[fig:box_1]{1}, we have to employ a suitable code for the logical auxiliary qubits. Here, we choose three $[[8, 3, 2]]$ codes. The red, blue and green cell of the tetrahedron are each mapped to one instance of the $[[8, 3, 2]]$ code. This code code shares the Z-plaquettes of the target $[[15, 1, 3]]$ code and it is also a $+1$-eigenstate of the weight-8 X-volume operators defined on the cells of the subsystem and the initial code and can therefore be used to extract the target stabilizers.

\section{Construction of feedback operation}\label{app:LUT_switching}

For switching between two codes we can construct the quantum feedback operation in an iterative manner, which is different to the approach for QEC. We first consider switching from the $[[15, 1, 3]]$ to the $[[7, 1, 3]]$ code. If a qubit storing a certain syndrome bit is in the $|1\rangle$ state, a certain Pauli-plaquette has to be applied: if the qubit storing the syndrome bit $A_X^R$ is in $|1\rangle$, we apply the Z-face $B_Z^{BG}$, if $A_X^B$ is in $|1\rangle$, we apply the Z-face $B_Z^{RG}$ and $A_X^G$ is in $|1\rangle$, we apply the Z-face $B_Z^{RB}$. If several syndrome bits are in the $|1\rangle$-state, the combination of the Pauli-plaquettes is applied which effectively flips some data qubits twice, thus implementing an identity operation on a subset of physical qubits. 
Note that in practice we only apply the respective operations on those data qubits which encode the target $[[7, 1, 3]]$ code and leave the qubits forming the yellow cell untouched. Before switching back, we re-initialize the yellow cell using the circuit shown in Fig.~\ref{fig:MF_init_aux}a. For the inverse switching direction, we use the same table but interchange the right and left column, e.g. if the syndrome bit storing the syndrome bit $B_Z^{BG}$ is in $|1\rangle$, we apply the X-face $A_X^R$. 

\begin{table}[!tb]
    \centering
    \renewcommand*{\arraystretch}{1.7}
    \begin{tabular}{|c|c|}
    \hline
     syndrome ($A_X^R, A_X^G, A_X^B$) & switching operation \\
     \hline
     (1, 0, 0)& $Z_2 Z_5 Z_{11} Z_{13}$\\
     \hline
     (0, 1, 0)& $Z_2 Z_3 Z_{12} Z_{13}$\\
     \hline
     (0, 0, 1)& $Z_1 Z_2 Z_{13} Z_{14}$\\
     \hline
     (1, 1, 0)& $Z_3 Z_5 Z_{11} Z_{12}$\\
     \hline
     (1, 0, 1) & $Z_1 Z_5 Z_{11} Z_{14}$\\
     \hline
     (0, 1, 1)& $Z_1 Z_3 Z_{12} Z_{14}$\\
     \hline
     (1, 1, 1)& $Z_1 Z_2 Z_3 Z_5 Z_{11} Z_{12} Z_{13} Z_{14}$\\
     \hline
     \hline
     syndrome ($B_Z^{BG}, B_Z^{RB}, B_Z^{RG}$) & switching operation \\
     \hline
     (1, 0, 0)& $X_0 X_1 X_2 X_3$\\
     \hline
     (0, 1, 0)& $X_1 X_2 X_4 X_5$\\
     \hline
     (0, 0, 1)& $X_2 X_3 X_5 X_6$\\
     \hline
     (1, 1, 0)& $X_0 X_3 X_4 X_5$\\
     \hline
     (1, 0, 1) & $X_0 X_1 X_5 X_6$\\
     \hline
     (0, 1, 1)& $X_1 X_3 X_4 X_6$\\
     \hline
     (1, 1, 1)& $X_0 X_2 X_4 X_6$\\
     \hline
    \end{tabular}
    \caption{\justifying \textbf{Feedback operation for switching between $\bm{[[15, 1, 3]]}$ and $\bm{[[7, 1, 3]]}$. } For each possible switching syndrome indicated in the left column, we apply the respective Pauli-operation on the right. }
    \label{tab:LUT_switching}
\end{table}

\section{Numerical methods}\label{app:numerical_methods}

We perform Monte Carlo simulations to determine the logical failure rates of our protocols~\cite{pecos_git}. Every component in the circuit is implemented by, first, applying the ideal operation followed by an error $E$ with a given probability $p$. Specifically, we implement a depolarizing channel after each single- and two-qubit gate. With probabilities $p_1$ and $p_2$ one of the errors in the sets $E_1$ and $E_2$ is applied and we can define the error channels as
\begin{align}
    \mathcal{E}_1(\rho) &= (1 - p_1)\rho + \frac{p_1}{3} \sum_{i= 1}^3 E^{i}_1 \rho E^{i}_1 \\
    \mathcal{E}_2(\rho) &= (1 - p_2)\rho + \frac{p_2}{15} \sum_{i= 1}^{15}   E_2^{i} \, \rho\, E_2^{i}.  \nonumber\label{eq:depol_single_qubit}
\end{align}
with $E_1^k \in \{X$, $Y$, $Z\}$ for $k = 1, 2, 3$ and 
$E_2^k$ $\in$ $\{IX$, $XI$, $XX$, $IY$, $YI$, $YY$, $IZ$, $ZI$, $ZZ$, $XY$, $YX$, $XZ$, $ZX$, $YZ$, $ZY\}$ for $k = 1, ..., 15$. 
Furthermore, we initialize and measure all qubits in the Z-basis and simulate faults on these components by applying X-flips after and before the respective operation with a given probability $p_{\mathrm{init}}$ and $p_{\mathrm{meas}}$. Additionally, idling qubits may dephase during measurements, which we model with the error channel 
\begin{align}
    \mathcal{E}_{\mathrm{idle, meas}}(\rho) &= (1 - p_{\mathrm{idle, meas}})\rho + p_{\mathrm{idle}} Z\rho Z. 
\end{align}
We simulate a simple single-parameter noise model where $p \coloneq p_2 = 10\cdot p_{\mathrm{init}} = 10 \cdot p_{\mathrm{meas}} = 10\cdot p_1$ and $p_{\mathrm{idle, meas}} = 0$ for the results shown in main text Fig.~\ref{fig:results_main}a. We include dephasing during measurements for the comparison to the measurement-based protocol as indicated on the y-axis in main text Fig.~\ref{fig:results_main}b.

Furthermore, we decompose the Toffoli-gates into single- and two-qubit gates, as shown in main text Fig.~\ref{fig:results_concatenated_code}. If at least one of the two control-qubits is not used afterwards, we do not need to apply the last 6 components shown in grey. We simulate the decomposed Toffoli-gate which we use in our protocol for all eight possible binary input states and determine the probability of the target qubit being flipped. We perform a linear fit for each input state and average the obtained slope. 
For error probabilities $p \coloneq p_2 = 10 p_1$ on single- and two-qubit gates, we find 
\begin{align}
    p_{\mathrm{toff}} = 2.88(10) \cdot p. \label{eq:toffoli_error_rate}
\end{align}
Analogously, we also determine the probability of flipping the respective target-qubit for two consecutive Toffoli-gates that share one control and the target-qubit, as used in parts of our MF FT switching protocols. In this case, we find
\begin{align}
    p_{\mathrm{2 toffs}} = 5.12(23) \cdot p < 2 \cdot p_{\mathrm{toff}}. 
\end{align}
We therefore simulate errors on each Toffoli-gate by flipping the target qubit with probability $p_{\mathrm{toff}}$, which slightly overestimates the total Toffoli-gate error rate. \\

\section{Effective error polynomial and fault-path counting}\label{app:effective_polynomial}

\begin{figure*}[!tb]
	\centering
	\includegraphics[width=0.66\linewidth]{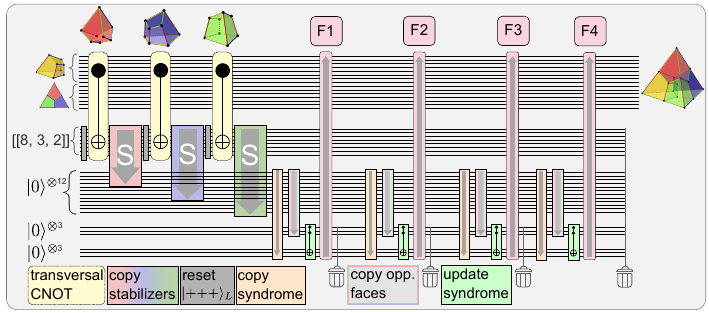}
	\caption{\justifying \textbf{Schematic circuit for MF FT switching from the $\bm{[[7, 1, 3]]}$ to the $\bm{[[15, 1, 3]]}$ code. }We employ a similar strategy as for the inverse switching direction: First, we initialize a logical auxiliary qubit in the $|+$$++\rangle_L$-state of the $[[8, 3, 2]]$ code (grey). Then, we couple those qubits belonging to the red cell to this encoded auxiliary register with a transversal CNOT-gate (yellow) and copy pairs of opposing X-faces of this cell to a register of physical qubits, which are initially prepared in $|0\rangle$ (pink). We repeat this procedure for the blue and green cell (blue, green). Then, the syndrome information is transferred to a set of physical qubits (orange), as well as the opposing faces belonging to the same cell (grey). Finally, the agreement of opposing faces within the same cell is checked by updating the extracted syndrome accordingly with a Toffoli-gate (green). The quantum feedback operation (F1-F4) is implemented in four steps with a reset of the updated syndrome in between (pink). }\label{fig:high_level_circuit_scheme_7_to_15}
\end{figure*}

We estimate the leading order contributions to the logical failure rate of a protocol, such as switching or state initialization, for the initial, non-concatenated code. For small physical error rates, these are
\begin{align}
    p_L &= c_2 p_2^2 + c_1 p_1^2 + c_{\mathrm{toff}} p_{\mathrm{toff}}^2 + c_{\mathrm{init}} p_{\mathrm{init}}^2 + c_{1, 2} p_1 p_2 \nonumber \\
    &+  c_{2, \mathrm{toff}} p_2 p_{\mathrm{toff}} + c_{2, \mathrm{init}} p_2 p_{\mathrm{init}} + c_{1, \mathrm{toff}} p_1 p_{\mathrm{toff}} \nonumber \\
    &+ c_{1, \mathrm{init}} p_1 p_{\mathrm{init}} + c_{\mathrm{init, toff}} p_{\mathrm{init}} p_{\mathrm{toff}} + \mathcal{O}(p^3),
    \label{eq:error_polynomial_full}
\end{align}
where the coefficients $c_{i, j}$ with $i, j = \{1, 2, \mathrm{toff}, \mathrm{init}\}$ correspond to the number of weight-2 faults on the specified components (single- and two-qubit gates, Toffoli-gates, physical qubit initializations) that lead to a logical failure and $p_1, p_2, p_{\mathrm{toff}}, p_{\mathrm{init}}$ are the error rates on the respective circuit components. 
However, the number of two-qubit gates in our MF FT code switching protocols is orders of magnitude larger than that of single-qubit gates and physical qubit initializations. Therefore, we estimate $c_1, c_{\mathrm{init}}, c_{1, \mathrm{init}}, c_{1, \mathrm{toff}}, c_{\mathrm{init}, \mathrm{toff}} \ll c_2$ and in the following neglect these coefficients in the above error polynomial. Furthermore, we consider smaller error rates on single-qubit gates and initializations than on two-qubit gates~\cite{postler2023demonstration} and approximate $p_1 = p_{\mathrm{init}} = \frac{p_2}{10}$ in our simulations. In this regime, we therefore also approximate that the contributions from error configurations with one fault on a two-qubit gate and another fault on a single-qubit gate or a physical qubit initialization scaling with $p_1$ and $p_{\mathrm{init}}$ are negligible. With these approximations, we find in leading order
\begin{align}
    p_L \approx c_2 p_2^2 +  c_{2, \mathrm{toff}} p_2 p_{\mathrm{toff}} + c_{\mathrm{toff}} p_{\mathrm{toff}}^2 + \mathcal{O}(p^3). 
    \label{eq:error_polynomial_methods}
\end{align}

For the switching and state initialization protocols we determine the coefficients in the error polynomial by deterministically placing all possible weight-2 configurations on the specified type of component, summarized in Tab.~\ref{tab:coefficients_FP_counting}. 

\begin{table}[!tb]
    \centering
    \renewcommand*{\arraystretch}{1.7}
    \begin{tabular}{|c|c|c|}
    \hline
     Protocol & coefficient & \# weight-2 faults\\
      & & $|+\rangle_L/|0\rangle_L$\\
     \hline
     & $c_2^{(\rightarrow)}$ & 300.05/386.60\\
     $[[15, 1, 3]] \rightarrow [[7, 1, 3]]$  & $c_{2, \mathrm{toff}}^{(\rightarrow)}$ & 69.06/18.94\\
     & $c_{\mathrm{toff}}^{(\rightarrow)}$  & 9/0\\
     \hline
     & $c_2^{(\leftarrow)}$& 1832.53/490.38\\
     $[[15, 1, 3]] \leftarrow [[7, 1, 3]] $& $c_{2, \mathrm{toff}}^{(\leftarrow)}$ & 178.67/469.33\\
     & $c_{\mathrm{toff}}^{(\leftarrow)}$ & 3/1220\\
    \hline 
    \raisebox{-8pt}[0pt][0pt]{$|0\rangle_L^{[[7, 1, 3]]}$} &$c_2^{(\mathrm{init})}$ & 27.88\\
    &$c_{2, \mathrm{toff}}^{(\mathrm{init})}$ & 8.27\\
    %\hline
    %TV CNOT $[[7, 1, 3]]$ & %$c_2^{(\mathrm{CNOT})}$ & 10.45\\
    \hline
    \end{tabular}
    \caption{\justifying \textbf{Coefficients from fault-path counting for different protocols. } We determine the number of weight-2 error configurations that lead to a logical failure for two input states $|+\rangle_L$ (left) and $|0\rangle_L$ (right) for each block, indicated by the two numbers in the right column. The index ``2'' means that two faults are placed on different two-qubit gates, the index ``2, toff'' means that one fault is placed on a two-qubit gate and one fault on a Toffoli-gate and ``toff'' indicates that two faults are placed on different Toffoli-gates. We divide the obtained number of logical failures by 15 if an error is placed on a two-qubit gate. Note that we do not yet take the effective Toffoli-gate error rate, given in Eq.~\eqref{eq:toffoli_error_rate} into account. }
    \label{tab:coefficients_FP_counting}
\end{table}

We plot Eq.~\eqref{eq:error_polynomial_methods} with the determined coefficients given in Tab.~\ref{tab:coefficients_FP_counting} for each switching direction and compare it to the logical failure rate obtained from Monte Carlo simulations, which is shown in main text Fig.~\ref{fig:results_main}a. 
We find that the logical failure rate and the determined polynomial agree within 6\% below $p = 10^{-3}$.

\section{Concatenated error polynomial}\label{app:concatenated_error_üplynomial}

\begin{figure*}[!tb]
	\centering
	\includegraphics[width=\linewidth]{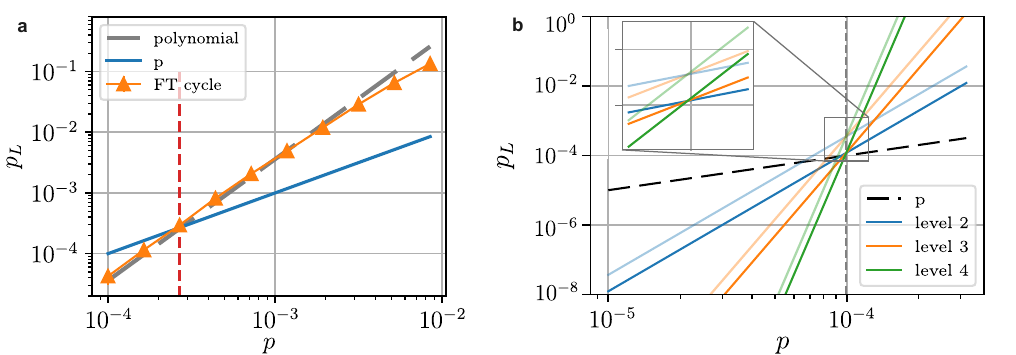}
	\caption{\justifying \textbf{MF FT code switching cycle and approximated scaling of logical T-gate. } \textbf{a} Logical error rate for a full switching cycle, starting and ending in the $[[7, 1, 3]]$ code. The grey dashed line corresponds to the summed polynomials of each individual switching step. We find a break-even point at approximately $p_{\mathrm{th}} = 2.6\cdot 10^{-4}$ as indicated by the red dashed line. \textbf{b} Approximated error polynomials of the logical error rates for the logical T-gate (darker colored) and the logical Toffoli-gate (lighter colored) for concatenation levels $l = 2, 3, 4$, based on the coefficients summarized in Tab.~\ref{tab:coefficients_FP_counting} and the polynomials given in Tab.~\ref{tab:summary_concatenated_rates} for small physical error rates $p$. The level-l Toffoli-gate error rates dominate the logical failure rates of the logical level-($l-1$) T-gate, and we find a pseudothreshold of $p_{th} \approx 1 \cdot 10^{-4}$ for both logical operations (grey dashed vertical line). }
	\label{fig:methods_full_switching_cycle}
\end{figure*}

Next, we extend this approach to the building blocks required for FT quantum computing, namely the initialization of a logical qubit in the $[[7,1,3]]$ code, final projective measurements, the logical gates H$_L$, T$_L$, CNOT$_L$ and Toffoli-gates on the seven-qubit color code, as well as switching in both directions and rounds of error correction, which have to be performed between logical operations in an algorithm~\cite{gottesman2013fault, aliferis2005quantum, cross2007comparative, gutierrez2015comparison, chamberland2016thresholds, zhou2024algorithmic}. Table~\ref{tab:summary_concatenated_rates} summarizes the dominating leading order contributions to the logical failure rate for each block in the column labeled ``level~1''. 

The logical failure rates for operations that have a natively transversal implementation on the $[[7, 1, 3]]$ code,~i.e.~projective measurements, the H- and the CNOT-gate, generalize in a straight-forward way to the next concatenation level: for 7 physically executed gates, there are maximally 
$\binom{7}{2}$ non-correctable error configuration occurring with a probability $p^{(0)2}$, where $p^{(0)}$ is the physical gate error rate. Analogously for concatenation level~$l+1$, the final logical error rate is given by $\binom{7}{2} \cdot p^{(l)2}$ with the failure probability $p^{(l)}$ on the next lower concatenation level~\cite{chamberland2016thresholds, paetznick2011fault}. 
Furthermore, we find that faults on CNOT- and Toffoli-gates dominate the total rate $p_L$ for switching between codes on the first level of concatenation while contributions from faulty single-qubit gate operations are negligible (main text Fig.~\ref{fig:results_main}a). 
We follow the same strategy to estimate the logical error rates of the remaining building blocks of qubit initialization and QEC: The total logical error rate is dominated by the two-qubit and Toffoli-gate error rates and we neglect the remaining parts of the polynomial. Note that the coefficient $c_{\mathrm{toff}}^{(\mathrm{init})} = 0$ for logical qubit initialization since is only one Toffoli-gate in the respective circuit. The distance of the concatenated code in this construction is given by
\begin{align}
    d' = \frac{1}{2} \cdot \left( d_1 d_2 + d_1 + d_2 - 1\right)\label{eq:distance}. 
\end{align}

Comparing the different logical error rates of the first level of concatenation, we see that the logical error rate of any logical operation that includes a switching step is dominated by this switching procedure, since this has a much higher error rate than the transversal CNOT- or single-qubit gates. Specifically, we need to switch in both directions once in order to implement the logical T-gate on the seven-qubit color code. The logical error rate of this operation T$_L^{(1)}$ is therefore approximately given by the error rates of the individual switching steps $p_{\leftarrow + \rightarrow}^{(1)} \approx p_{\leftarrow}^{(1)} + p_{\rightarrow}^{(1)} $. This approximation holds if one switching direction has much larger coefficients in the error polynomial than the other direction, which is the case here as summarized in Tab.~\ref{tab:coefficients_FP_counting}. Additionally, we verify this approximation by simulating a complete cycle of switching and compare this to the summed polynomials of each direction, as shown in Fig.~\ref{fig:methods_full_switching_cycle}a. 

\begin{table*}[!tb]
    \centering
    \renewcommand*{\arraystretch}{1.9}
    \begin{tabular}{|c|c|c|c|}
    \hline
     Operation & physical & level 1 & level $l+1$\\
     \hline
     Initialization $|0\rangle$ & $\frac{p_2^{(0)}}{10}$& $c_{\mathrm{2}}^{(\mathrm{init})} p_2^{(0)2} + c_{\mathrm{2, toff}}^{(\mathrm{init})} p_2^{(0)} p_{\mathrm{toff}}^{(0)} + c_{\mathrm{init, toff}}^{(\mathrm{init})} \frac{p_2^{(0)}}{10} p_{\mathrm{toff}}^{(0)}$  & $\approx c_{\mathrm{init, toff}}^{(\mathrm{init})} p_{\mathrm{init}}^{(l)} p_{\mathrm{toff}}^{(l)}$\\
     \hline
     Measurement & $\frac{p_2^{(0)}}{10}$ & $\binom{7}{2} (\frac{p_2^{(0)}}{10})^2$ & $\binom{7}{2} (\frac{p_2^{(l)}}{10})^2 \ll p_{\mathrm{toff}}^{(l)2}$\\
     \hline
     Single-qubit H-gate & $\frac{p_2^{(0)}}{10}$ & $\binom{7}{2} (\frac{p_2^{(0)}}{10})^2$ & $\binom{7}{2} (\frac{p_2^{(l)}}{10})^2 \ll p_{\mathrm{toff}}^{(l)2}$\\
     \hline
     Single-qubit T-gate & $\frac{p_2^{(0)}}{10}$  & $\approx p_{\leftarrow}^{(1)} + p_{\rightarrow}^{(1)}$ & $ \approx (c_{\mathrm{toff}}^{(\leftarrow)} + c_{\mathrm{toff}}^{(\rightarrow)}) \cdot p_{\mathrm{toff}}^{(l)2}$\\
     \hline
     CNOT-gate & $p_2^{(0)}$ & $\binom{7}{2} p_2^{(0)2}$& $\binom{7}{2} p_2^{(l)2} \ll p_{\mathrm{toff}}^{(l)2}$\\
     \hline
     Reduced Toffoli-gate & $p_{\mathrm{toff, red}}^{(0)}$ &$ \approx 3 p_{\leftarrow}^{(1)} + p_{\rightarrow}^{(1)}$ & $ \approx (3c_{\mathrm{toff}}^{(\leftarrow)} + c_{\mathrm{toff}}^{(\rightarrow)}) \cdot p_{\mathrm{toff}}^{(l)2}$ \\
     \hline
     Toffoli-gate & $p_{\mathrm{toff} }^{(0)}$ &$\approx 3 p_{\leftarrow}^{(1)} + 3 p_{\rightarrow}^{(1)}$ &  $ \approx 3(c_{\mathrm{toff}}^{(\leftarrow)} + c_{\mathrm{toff}}^{(\rightarrow)}) \cdot p_{\mathrm{toff}}^{(l)2}$\\
     \hline
    $[[15, 1, 3]] \rightarrow [[7, 1, 3]]$& - & $p_{\rightarrow}^{(1)} = c_2^{(\rightarrow)} p_2^{(0)2} +  c_{2, \mathrm{toff}}^{(\rightarrow)} p_2^{(0)} p_{\mathrm{toff}}^{(0)} + c_{\mathrm{toff}}^{(\rightarrow)} \cdot p_{\mathrm{toff}}^{(0)2}$ & $\approx c_{\mathrm{toff}}^{(\rightarrow)} \cdot p_{\mathrm{toff}}^{(l)2}$\\
     \hline
    $[[15, 1, 3]] \leftarrow [[7, 1, 3]]$& - & $p_{\leftarrow}^{(1)} = c_2^{(\leftarrow)} p_2^{(0)2} +  c_{2, \mathrm{toff}}^{(\leftarrow)} p_2^{(0)} p_{\mathrm{toff}}^{(0)} + c_{\mathrm{toff}}^{(\leftarrow)} \cdot p_{\mathrm{toff}}^{(0)2}$ & $\approx c_{\mathrm{toff}}^{(\leftarrow)} \cdot p_{\mathrm{toff}}^{(l)2}$\\
     \hline
     QEC round & - & $p_{\mathrm{QEC}}^{(1)} = c_2^{(\mathrm{QEC})} p_2^{(0)2} +  c_{2, \mathrm{toff}}^{(\mathrm{QEC})} p_2^{(0)} p_{\mathrm{toff}}^{(0)} + c_{\mathrm{toff}}^{(\mathrm{QEC})} \cdot p_{\mathrm{toff}}^{(0)2}$ & $\approx c_{\mathrm{toff}}^{(\mathrm{QEC})} \cdot p_{\mathrm{toff}}^{(l)2}$\\
     \hline
    \end{tabular}
    \caption{\justifying \textbf{Approximated error rates for operations on the concatenated code. } For each operation indicated on the left, we estimate the leading order contributions to the logical failure rates for concatenation level~$l$. The error rates for the non-concatenated case (second column) correspond to the physical error rates on each component. Each physical qubit is then replaced by a $[[7, 1, 3]]$ logical qubit in concatenation level 1 (third column). Here, the indicated superscript of the coefficient $c$ specifies the considered building block and the subscript specifies the considered component,~e.g. $c_{\mathrm{2}}^{(\mathrm{init})}$ is the number of faults only on two-qubit gates during the (logical) qubit initialization. The level of concatenation is indicated by the superscript on the error rates, as for example $p_2^{(0)}$ is the error rate of the physical two-qubit gates. We determine the specified coefficients with fault-path counting, which are explicitly given in Tab.~\ref{tab:coefficients_FP_counting}. We include the approximation discussed in the main text Sec.~\ref{sec:scalability} and neglect parts of the full polynomial. The most costly component in our circuits is the concatenated Toffoli-gate, because it includes several switching operations, each with a large gate overhead. We therefore find that for high levels of concatenation, the logical error rates of the non-transversal logical operations is dominated by the effective Toffoli-gate error rate $p_{\mathrm{toff}}^{(l)} \approx (3[c_{\mathrm{toff}}^{\rightarrow} + c_{\mathrm{toff}}^{\leftarrow}])^{2^l-1} \cdot p_{\mathrm{toff}}^{(1)2^l}$. }\label{tab:summary_concatenated_rates}
\end{table*}

\section{Scaling of the logical failure rate for concatenated scheme}\label{app:concatenated_code_switching}

The Toffoli-gate on the first level of concatenation also contains switching steps as illustrated in main text Fig.~\ref{fig:results_concatenated_code}. We consider two versions of the Toffoli-gate: the reduced Toffoli-gate, where the control-qubits are not used afterwards and we only apply the operations shown in black, and the full Toffoli-gate, which includes the full decomposition and we also execute the operations depicted in grey. If each qubit is an encoded $[[7, 1, 3]]$ code before applying the Toffoli-gate, we have to switch to the $[[15, 1, 3]]$ code after the first H$_L$-gate in order to apply the logical T$_L$ and the transversal CNOT$_L$-gates and, again, switch back to the $[[7, 1, 3]]$ code in the end before the final H$_L$-gate. The logical error rate for the Toffoli-gate is therefore also dominated by the switching procedure and can be approximated by summing up the individual switching error rates. Note that summing up all individual switching error rates on control- and target-qubits of the logical Toffoli-gate overestimates the final probability of flipping only the respective target qubit. 

We now determine the dominant contributions to the logical failure rate for concatenation level~$l+1$ for the not inherently transversal logical gates of the $[[7, 1, 3]]$ code. The rate of the full Toffoli-gate on level~2 is again dominated by the switching procedure,~i.e. the probabilities $p_{\rightarrow}^{(2)}$ and $p_{\leftarrow}^{(2)}$, which in turn are dominated by the logical Toffoli-gate of level~1. Iterating this to level~$l+1$, we find that the logical operations of concatenation level~$l+1$ will always be dominated by the Toffoli-gate error rate, since this includes the most switching steps and, therefore the noisiest parts of the respective logical operation. Fig.~\ref{fig:methods_full_switching_cycle}b shows the resulting polynomial with the leading order contributions to the logical failure rate, as given in the third column of Tab.~\ref{tab:summary_concatenated_rates}, for the logical concatenated T$_L$- and the Toffoli-gate. We find a similar pseudothreshold at $p_{th} \approx 1 \cdot 10^{-4}$ for both protocols, since the noisy Toffoli-gates dominate the effective total logical error rate.

\section{Cost of QEC rounds}\label{app:cost_of_QEC}

Like the MF FT code switching protocols, also the QEC blocks of concatenation level $l+1$ are dominated by the logical Toffoli-gate error rate, as this presents the component with the largest error rate. Exemplarily considering protocol~\cite{veroni2024optimized}, we find an upper bound of the dominant coefficient $c_{\mathrm{toff}}^{(\mathrm{QEC})}$ of the logical failure rate for one round of QEC to be
\begin{align}
    c_{\mathrm{toff}}^{(\mathrm{QEC})} \leq \binom{\#\mathrm{Toffoli-gates}}{2} =  \binom{21}{2} \ll c_{\mathrm{toff}}^{(\leftarrow)} +  c_{\mathrm{toff}}^{(\rightarrow)}. 
\end{align}
The cost of one round of QEC below the pseudothreshold for concatenation level~$l > 1$ is therefore much smaller than the cost of a logical T-gate on the concatenated $[[7, 1, 3]]$ code. 

\begin{table}[!tb]
    \centering
    \renewcommand*{\arraystretch}{1.7}
    \begin{tabular}{|c|c|c|c|}
    \hline
     Protocol & CNOTs & Toffolis & qubits\\
     \hline
     $[[15, 1, 3]] \rightarrow [[7, 1, 3]]$ & 120 & 8 & 29/53\\
     \hline
     $[[15, 1, 3]] \leftarrow [[7, 1, 3]] $ & 296 & 40 & 35/131\\
    \hline
     $|0\rangle_L^{[[7, 1, 3]]}$ & 15 & 1 & 9\\
    \hline
      $|+++\rangle_L^{[[8, 3, 2]]}$ & 32 & 4& 12/16\\
    \hline
     T$_{[[7, 1, 3]]}$ & 416 & 48 & 35/169\\
    \hline
    \end{tabular}
    \caption{\justifying \textbf{Resources required for MF FT building blocks. } Number of CNOT- and Toffoli-gates and the number of physical qubits required for the protocols indicated on the left. This includes MF switching, the initialization of the logical auxiliary qubits and the logical T-gate on the 2D color code. The two given numbers for the qubit count correspond to the protocol with qubit resets (left) and without qubit resets (right).}\label{tab:ressources_MF_operations}
\end{table}

\section{MF FT initialization of logical auxiliary qubits}\label{app:MF_logical_init}
 
\begin{figure*}[!tb]
	\centering
	\includegraphics[width=\linewidth]{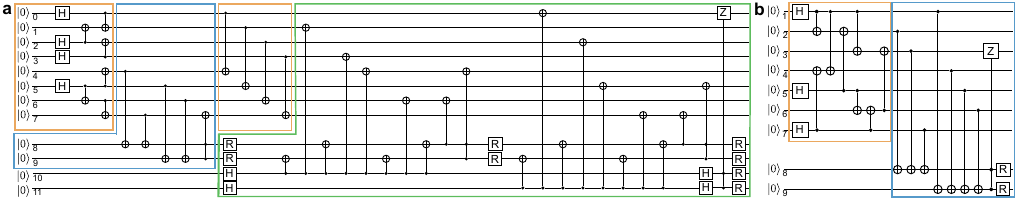}
	\caption{\justifying \textbf{Full circuit for MF FT initialization of logical auxiliary qubits. }\textbf{a} The $|+$$++\rangle_L$-state of the $[[8, 3, 2]]$ code can be initialized by, first, preparing two GHZ-states (left orange box) and entangling these (right orange box). We perform a verification before entangling the two GHZ-states (blue) in order to detect potentially dangerous weight-2 X-errors. Finally, we need to verify that no weight-2 Z-error, which could result from a single fault in the circuit, is present on the qubit register. To this end, we extract two complementary logical X-operators which are chosen such that they anticommute with all potentially dangerous Z-error configurations (green box). \textbf{b} The $|0\rangle_L$ on the $[[7, 1, 3]]$ code can be initialized MF in a similar way by, first, implementing a non-FT encoding (orange), and then mapping suitable operators onto two physical auxiliary qubits and applying a Toffoli-gate (blue)~\cite{heussen2024measurement, veroni2024optimized}. } 
	\label{fig:MF_init_aux}
\end{figure*}

To ensure fault tolerance for switching from $[[7, 1, 3]]$ to $[[15, 1, 3]]$ we have to verify that no weight-2 X-error and no weight-2 Z-error is present on the encoded logical auxiliary state. In this switching direction, we make use of an auxiliary $[[8, 3, 2]]$, as specified in Fig.~\ref{fig:codes_definition}c. We construct a MF FT circuit for the initialization of the logical $[[8, 3, 2]]$ qubit in the $|+$$++\rangle_L$-state by extending the non-FT encoding~\cite{wang2024fault} and adding a verification. First, we prepare two weight-4 GHZ-states on separate sets of qubits and entangle these for the non-FT encoding. 
A weight-two Z-error is equivalent to a logical Z$_L$ on the auxiliary qubit and would propagate onto two data qubits when the auxiliary and data register are coupled with the transversal CNOT-gate, as illustrated in Fig.~\ref{fig:high_level_circuit_scheme_7_to_15}. We can correct this weight-2 Z-error by mapping two complementary logical X$_L$ operators to the auxiliary system by building on a flag-qubit based scheme~\cite{veroni2024optimized}, as shown in Fig.~\ref{fig:MF_init_aux}a in the green box. The two logical operators are chosen such that only the dangerous weight-2 Z-errors anticommute with both of them. In the end we apply a Toffoli-type feedback operation so that a correction is only applied if both extracted operators are in the $|1\rangle$-state. 
Furthermore, a weight-2 X-error on the auxiliary register would propagate onto the second auxiliary register storing the syndrome information bits, similarly to X-errors on data qubits. We find that there are only two inequivalent weight-2 error configurations that can result from a single fault in the non-FT encoding by placing a fault at every possible location in the circuit. We can align one of these error configurations such that it flips the extracted syndrome bits either exactly twice or not at all. The second dangerous error configuration is corrected by extracting the instantaneous stabilizers to an auxiliary system and applying a Toffoli-type feedback operation (blue box in Fig.~\ref{fig:MF_init_aux}a) before entangling the two separate GHZ-states. 

\section{Resources and circuits}\label{app:resources_and_circuits}

Table~\ref{tab:ressources_MF_operations} summarizes the required resources for the constructed MF FT protocols in terms of qubit count, number of two- and three-qubit gates.

\begin{figure*}[b]
	\centering\includegraphics[width=\linewidth]{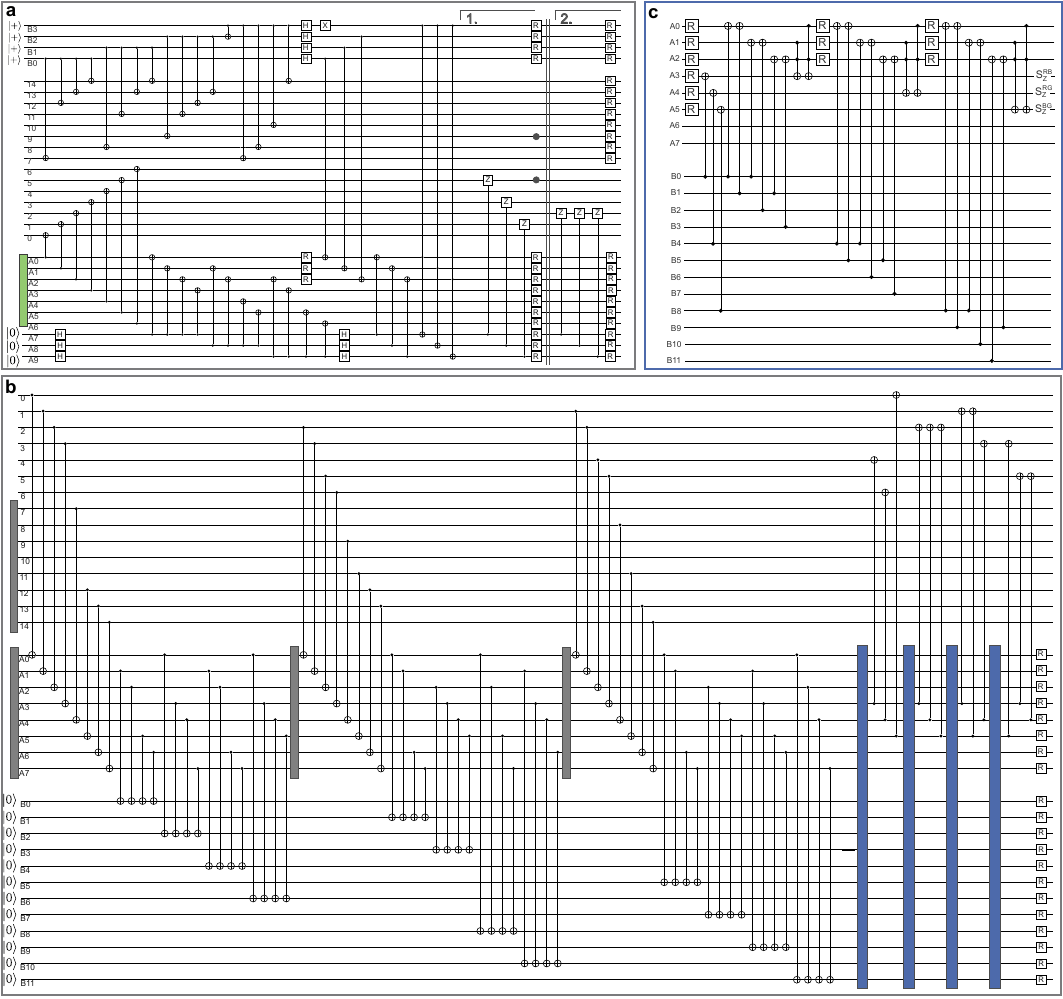}
	\caption{\justifying \textbf{Full circuits for MF FT switching between the $\bm{[[15, 1, 3]]}$ and the $\bm{[[7, 1, 3]]}$ code. } \textbf{a} Circuit for switching from $[[15, 1, 3]]$ to $[[7, 1, 3]]$. First, all shown gates up to the double-line (fourth to last position) are executed. Then, we start from the beginning and, again, perform all gates up to bracket number 1 and jump into the second bracket 2 to execute the last three CZ-gates. The operation $R$ refers to qubit \textit{reset}, which is either done by reinitializing the physical qubit in the $|0\rangle$ state or replacing it with a fresh qubit. The green box corresponds to the MF encoding of the logical $|0\rangle_L$ state on the $[[7, 1, 3]]$ color code, as given in~\cite{heussen2024measurement} and shown in Fig.~\ref{fig:MF_init_aux}b. \textbf{b} Circuit for switching from $[[7, 1, 3]]$ to $[[15, 1, 3]]$. The grey boxes correspond to the MF initialization of the logical auxiliary state $|+$$++\rangle_L$ shown in Fig.~\ref{fig:MF_init_aux}a and the blue boxes correspond to the syndrome update shown in c. \textbf{c} Circuit for updating the syndrome based on agreement check. Qubits A0, A1 and A2 are used to check the agreement of opposing faces by copying both respective syndrome bits onto the same qubit in register A. Qubits A4, A5 and A6 store the syndrome bit information, which is updated with two Toffoli-gates each. Using two successive Toffoli-gates is still FT because they only share one control-qubit and, therefore, no single fault in the first Toffoli-gate can cause an erroneous flip on the second one. }
	\label{fig:collected_circuits}
\end{figure*}

Figure~\ref{fig:collected_circuits} shows the explicit circuits that we implement for switching between the tetrahedral $[[15, 1, 3]]$ code and the $[[7, 1, 3]]$ code. Here, the \textit{reset}-operation $R$ includes the reinitialization of the respective physical qubit or the substitution with a fresh auxiliary qubit in a pure state.

\section*{Acknowledgments}

We would like to thank Sebastian Weber for useful discussions.
We acknowledge funding by the Federal Ministry of Education and Research (BMBF) project MUNIQC-ATOMS (Grant No. 13N16070).
F.B., D.L. and M.M. additionally acknowledge support from the German Research Foundation (DFG) under Germany’s Excellence Strategy ‘Cluster of Excellence Matter and Light for Quantum Computing (ML4Q) EXC 2004/1’ 390534769,
the BMBF via the VDI within the project IQuAn,
the ERC Starting Grant QNets through Grant No. 804247, 
the US Army Research Office through Grant Number W911NF-21-1-0007, 
the Intelligence Advanced Research Projects Activity (IARPA) under the Entangled Logical Qubits program through Cooperative Agreement Number W911NF-23-2-0216, 
the European Union (EU) Horizon Europe research and innovation program under Grant Agreement No. 101114305 (“MILLENION-SGA1” EU Project),
and the Munich Quantum Valley (K-8), which is supported by the Bavarian state government with funds from the Hightech Agenda Bayern Plus.
H.P.B. and K.B. additionally acknowledge funding from the BMBF under the grant QRydDemo, and from the Horizon Europe program HORIZON-CL4-2021-DIGITAL- EMERGING-01-30 via the project 101070144 (EuRyQa).
We gratefully acknowledge the computing time provided at the NHR Center NHR4CES at RWTH Aachen University (Project No.~p0020074). This is funded by the Federal Ministry of Education and Research and the state governments participating based on the resolutions of the GWK for national high-performance computing at universities.

\section*{Data availability}
The data provided in the figures in this article and the explicit circuits are available at \url{https://doi.org/10.5281/zenodo.13941146}. The full simulation code can be provided by the corresponding author upon reasonable request.

\section*{Author contributions:}
F.B. developed the presented protocols and performed the numerical simulations. F.B., D.L., and K.B. analyzed results and wrote the manuscript, with contributions from all authors. H.P.B. and M.M. supervised the project.

\section*{Competing interests:}
The authors declare no competing interests.

\end{document}